\documentclass[12pt, a4paper]{iopart}
\usepackage{iopams}
\usepackage{setstack, cite,graphicx}
\begin{document}
\eqnobysec
\title[Localized Coherent Structures]{Localized Coherent Structures of Ishimori 
Equation I through Hirota's Bilinearization method:Time dependent/Stationary boundaries}
\author{S. Vijayalakshmi$^{\dag}$ and M. Lakshmanan$^{\dag \dag}$}
\address{$^{\dag}$Postgraduate and Research Department of Physics, Government Arts
College(Autonomous), Coimbatore-641 018, India}

\address{$^{\dag\dag}$Centre for Nonlinear Dynamics, Department of Physics, Bharathidasan
University, Tiruchirapalli-620 024, India}

\date{\today}

\begin{abstract} 
Ishimori equation is a $(2+1)$ dimensional generalization of the $(1+1)$
dimensional integrable classical continuous Heisenberg ferromagnetic spin
equation. The richness of the coherent structures admitted by Ishimori equation
I such as dromion, lump and rationally- exponentially localized solutions, have
been demonstrated in the literature through  $\bar \partial$ technique and
binary Darboux transformation method. To our knowledge Hirota's method had been
adopted to construct only the vortex solutions of Ishimori equation II. For the
first time, the various types of localized  coherent structures mentioned above
have been constructed in this  paper for the Ishimori equation I using the
Hirota's direct method. In particular we have  brought out the significance of
boundaries and arbitrary functions in generating all these types of localized structures and proved
that the absence of such boundaries leads only  to line soliton solutions. 
\end{abstract}

\maketitle

\section{Introduction}

A $(2+1)$ dimensional integrable generalization of the (1+1) dimensional integrable Heisenberg
ferromagnetic spin equation (isotropic Landau-Lifshitz equation) $\vec
S_t(x,t)=\vec S\wedge \vec S_{xx}$\cite{laks:1976} was introduced by Ishimori in
1984\cite{Ishi:1984} to explain the dynamics of the classical spin system on a
plane. Its form is
\numparts
\label{eqn1}
\begin{eqnarray}
\vec S_t(x,y,t)=\vec S\wedge (\vec S_{xx}+\sigma^2\vec S_{yy})+\phi_y\vec
S_x+\phi_x\vec S_y,\\
\phi_{xx}-\sigma^2\phi_{yy}=-2\sigma^2 \vec S\cdot \vec S_x \wedge \vec S_y,
\end{eqnarray}
\endnumparts
where $\vec S=(S_1, S_2, S_3)$ is the three dimensional spin unit vector $(\vec
S^2=1)$, $\phi (x,y,t)$ is a scalar field and $\sigma^2=\pm 1$. If $\sigma^2=+1$,
eq.~(1.1) is referred to as Ishimori equation I(IE I) and if $\sigma^2=-1$, it
is referred to as Ishimori equation II(IE II). An important feature associated
with eq.~(1.1) is the existence of nontrivial topological invariant known as
topological charge defined as 
\begin{equation}
Q=\frac{1}{4\pi}\int \int \vec S\cdot \vec S_x \wedge \vec S_y dx dy 
\label{eq2.2}
\end{equation}
and the solutions of eq.~(1.1) are classified in terms of the integer values of $Q$.
Just like the (1+1) dimensional integrable spin system is geometrically
equivalent\cite{laks:1976} to the nonlinear Schr\"odinger equation through a
moving space curve formalism, eq.~(1.1) is geometrically equivalent to the
Davey-Stewartson equation through a moving surface formalism\cite{laks:1998}. 

The initial value problems of both IE I and IE II have been analysed by the
$\bar \partial$ and nonlocal Riemann-Hilbert problem methods, respectively, in
\cite{BGK:1990} by Konopelchenko and Matkarimov. For stationary boundaries, the
initial boundary value problem for the IE I has been studied in \cite{BGK:1991}
and three different types of localized solutions (soliton-soliton(ss),
soliton-breather(sb), breather-breather(bb)) have been presented. The line
solitons admitted by both IE I and IE II have been presented in
\cite{BGK:1993}. The localized coherent structures for the  IE I have been
analysed in \cite{BGK:1992} for time dependent boundaries and the solutions 
such as rationally localized soliton, exponentially localized soliton and
rationally-exponentially localized soliton have been reported using inverse
scattering transform (IST) method. IE I has also been  analysed through binary Darboux
transformation method and different types of solutions have been constructed in
terms of grammian determinants in  \cite{IMAI}. Curiously, the Hirota's
bilinearization method, which is one of the celebrated direct methods and
applicable to almost all integrable soliton equations in (1+1) and (2+1)
dimensions, has not been applied so far (as far as our knowledge goes) to
obtain localized solutions for IE I, though it has been used by Ishimori
himself\cite{Ishi:1984} to obtain vortex solutions to IE II. The difficulty
probably lies in introducing the boundaries in the bilinearized form
appropriately for IE I.  In  this paper, we have successfully obtained all
types of localized structures of  IE I reported in
\cite{BGK:1991,BGK:1993,BGK:1992} through the Hirota's bilinearization
technique and presented the different types of localized structures admitted by
it for the case of both time dependent and stationary boundaries.

The paper is organised as follows: In section 2, eq.~(1.1) is bilinearized in
laboratory coordinates 
through stereographic projection and Painlev\'e property and multiline soliton solutions for both
IE I and IE II are presented for completeness. In section 3, the bilinearized version of IE I
 in terms of light cone coordinates is presented and the general
form of the solution is given. In particular, we point out how boundaries can
be introduced explicitly into the bilinearized version of eq.~(1.1). The role
of linear equations of modified Kadomtsev-Petviashvili in obtaining solutions of
the bilinear equations is also pointed out. 
The different types of localized coherent structures driven by time-dependent
boundaries are presented in section 4. In section 5, the behaviour of IE I in
the background of stationary boundaries is analysed. Importance of boundaries
in generating the various types of localized structures is also discussed.
Finally, the results are summarized in section 6, where the importance of
arbitrary function in expressing solutions is pointed out.

\section{Bilinearization of IE}

By making a stereographic projection of the spin of unit sphere on a complex
plane, the spin components can be written in terms of the stereographic
variable $\omega$\cite{laks:physica} as
\begin{eqnarray}
S^+=S_1+iS_2=\frac {2\omega}{1+|\omega|^2},\quad S_3=\frac {1-|\omega|^2}{1+|\omega|^2}
\end{eqnarray}
and eq.~(1.1) takes the form
\numparts
\begin{eqnarray}
i\omega_t+\omega_{xx}+\sigma^2\omega_{yy}-\frac {2\omega^*}
{1+|\omega|^2}(\omega_x^2+\sigma^2\omega_y^2)-i\phi_y\omega_x-i\phi_x\omega_y=0,\\
\phi_{xx}-\sigma^2\phi_{yy}=\frac
{4i\sigma^2}{(1+|\omega|^2)^2}(\omega_x^*\omega_y-\omega_x\omega_y^*).
\end{eqnarray}
\endnumparts
We find that this form is more convenient for further analysis as discussed
below. 

\subsection{Painlev\'e singularity structure analysis}

We can confirm the integrability nature of eq.~(2.2), by performing a Painlev\'e
analysis of it. Denoting $\omega$ and $\omega^*$ by $F$ and $G$ respectively, we
rewrite eq.~(2.2) as 
\numparts
\begin{eqnarray}
\fl (1+FG)
[iF_t+F_{xx}+\sigma^2F_{yy}-i(\phi_yF_x+\phi_xF_y)]-2G(F_x^2+\sigma^2F_y^2)=0,\\
\label{2.3a}
\fl (1+FG)
[-iG_t+G_{xx}+\sigma^2G_{yy}-i(\phi_yG_x+\phi_xG_y)]-2F(G_x^2+\sigma^2G_y^2)=0,\\
\label{2.3b}
\fl (1+FG)^2 (\phi_{xx}-\sigma^2\phi_{yy})=4i\sigma^2(F_yG_x-F_xG_y),
\label{2.3c}
\end{eqnarray}
\endnumparts
where eq.~(2.3b) is the complex conjugate of eq.~(2.3a). In order to carry out
a singularity structure analysis of eq.~(2.3), we effect the following local
Laurent expansion for each dependent variable in the neighbourhood of a
noncharacteristic singular manifold $\psi(x,y,t)=0$:
\begin{eqnarray}
\fl F=\psi^m\sum_{j=0}^{\infty}F_j(x,y,t)\psi^j,\;G=\psi^n\sum_{j=0}^{\infty} 
G_j(x,y,t)\psi^j,\;\phi=\psi^p\sum_{j=0}^{\infty} \phi_j(x,y,t)\psi^j.
\end{eqnarray}
We now substitute eq.~(2.4) into eq.~(2.3) and look at the leading order
behaviour of $\psi$. Here we come across two different branches, one at $m=0,
n=-1, p=0$ and another one at $m=-1, n=0, p=0$. In both the cases, $F_0$, $G_0$
and $\phi_0$ are found to be arbitrary functions of $x$, $y$ and $t$ and the
resonances are found to occur at $j=-1,0,0,0,1,1$. Further analysis confirms
that two arbitrary functions occur at the resonance values of $j=1$, without
the introduction of any movable critical manifold, while the remaining
coefficients can be expressed in terms of the earlier ones. Hence eq.~(2.2)
passes the Painlev\'e test and confirm its integrability nature.

\subsection{Bilinearization in laboratory coordinates}

To construct a formal B\"acklund transformation, we truncate the Laurent series
at the constant level term, that is (for the case $m=-1$, $n=0$ and $p=0$)
\begin{eqnarray}
F=F_0\psi^{-1}+F_1,\;\;\;\;G=G_0,\;\;\;\phi=\phi_0.
\end{eqnarray}
We can now construct the bilinear form of eq. (2.2), by considering
\begin{eqnarray}
F_1=0.\nonumber
\end{eqnarray}
Let $F_0=g$ and $\psi=f$, then $\omega=\frac{g}{f}$. Under this transformation
$\omega=\frac{g}{f}$,, where $g$ and $f$ are complex
functions of $x$, $y$ and $t$, eq. (2.2) can be written  in terms of the Hirota's
D-operators(which are defined as $D_x^iD_y^jD_t^ka(x,y,t)\cdot b(x,y,t)=(\partial_x-\partial_{x'})^i
(\partial_y-\partial_{y'})^j$\\
$(\partial_t-\partial_{t'})^k a(x,y,t)
b(x',y',t')|_{x=x',y=y',t=t'}$) as 
\numparts
\begin{eqnarray}
(iD_t-D_x^2-\sigma^2D_y^2)(f^*\cdot g)=0,\\
(iD_t-D_x^2-\sigma^2D_y^2)(f^*\cdot f-g^*\cdot g)=0,\\
\fl D_x(D_x(f^*\cdot f+g^*\cdot g))\cdot (f^*f+g^*g)=\sigma^2 
D_y(D_y(f^*\cdot f+g^*\cdot g))\cdot (f^*f+g^*g),
\end{eqnarray}
so that 
\begin{eqnarray}
\phi_x=-2i\sigma^2 \frac {D_y(f^*\cdot f+g^*\cdot g)} {(f^*f+g^*g)},\\
\phi_y=-2i \frac {D_x(f^*\cdot f+g^*\cdot g)} {(f^*f+g^*g)}.
\end{eqnarray}
\endnumparts
Note that eq.~(2.6c) arises due to the compatibility condition
$\phi_{xy}=\phi_{yx}$ and it is biquadratic.

\subsection{The line solitons}

Now, the construction of the line soliton solutions to IE becomes standard and
we briefly indicate their forms. One expands the
functions $g$ and $f$ as power series in the arbitrary parameter $\epsilon$ as
follows,
\begin{eqnarray}
g=\sum_{n=0}^{\infty} \epsilon^{2n+1}g_{2n+1},\quad f=1+\sum_{n=1}^{\infty}
\epsilon^{2n}f_{2n}.
\end{eqnarray}
Substituting these expansions into eqs. (2.6a)-(2.6c) and equating the
coefficients of various powers of $\epsilon$, we get the respective following 
system of equations from (2.6a), (2.6b) and (2.6c):
\numparts
\begin{eqnarray}
\fl \epsilon^{2n+1}:\;(i\partial_t+\partial_x^2+\sigma^2\partial_y^2) g_{2n+1}=-\sum_{k+m=n} D'
(f^*_{2k}\cdot g_{2m+1}),\\
\fl \epsilon^{2n}:\;\;i\partial_t (f^*_{2n}-f_{2n})-(\partial_x^2+\sigma^2\partial_y^2)
(f^*_{2n}+f_{2n})=\nonumber \\
\;\;D'\left (\sum_{n_1+n_2=n-1} (g^*_{2n_1+1}\cdot g_{2n_2+1})-\sum_{m_1+m_2=n}
(f^*_{2m_1}\cdot f_{2m_2})\right ),\\
\fl(\partial_x^2-\sigma^2\partial_y^2)
(f^*_{2n}-f_{2n})+D^{''}\left (\sum_{n_1+n_2=n-1} (f^*_{2n_1}\cdot f_{2n_2})+
\sum_{m_1+m_2=n-1}(g^*_{2m_1}\cdot g_{2m_2})\right )+\nonumber \\
\fl \left \{D_x(f^*_{(2n-2)x}-f_{(2n-2)x}+\sum_{n_1+n_2=n-2}D_x g^*_{2n_1+1}
\cdot g_{2n_2+1})-\sigma^2D_y (f^*_{(2n-2)y}-f_{(2n-2)y}+ \right.
\nonumber \\
\fl \left .\sum_{n_1+n_2=n-2}D_y
g^*_{2n_1+1}\cdot g_{2n_2+1}) \right \}\cdot (f^*_{2n-2}-f_{2n-2}+
\sum_{n_1+n_2=n-2}g^*_{2n_1+1}\cdot g_{2n_2+1})=0
\end{eqnarray}
\endnumparts
with $D'=iD_t-D_x^2-\sigma^2D_y^2$, $D^{''}=D_x^2-\sigma^2D_y^2$ and $f_0=0$. 
\subsubsection{1-soliton solution}
In order to construct the exact $N$-soliton solutions (N-SS) of eq.(1.1), 
we make the ansatz
\begin{eqnarray}
g_1=\sum_{j=1}^N \exp {\chi_j},\quad \chi_j=l_jx+m_jy+n_jt,
\end{eqnarray}
where $l_j$, $m_j$ and $n_j$ are complex constants. As an example, we write the
forms of $g$ and $f_2$ with the help of eq.~(2.8) for $N=1$ as
\begin{eqnarray}
g_1=M \exp {\chi_1},\quad \chi_1=l_1x+m_1y+n_1t,\quad f_2=\exp
{2(\chi_{1R}+\psi)},
\end{eqnarray}
where \\
$\chi_1=\chi_{1R}+i\chi_{1I}$, $n_1=i(l_1^2+\sigma^2 m_1^2)$ and $\exp {2\psi}=
\frac {\sigma^2 m_1^2-l_1^2}{(l_1+l_1^*)^2-\sigma^2(m_1+m_1^*)^2} MM^*$ and $M$
is an arbitrary complex constant. Now we distinguish the two cases $\sigma^2=+1$
and $\sigma^2=-1$\\ 
\underline{Case(i): $\sigma^2=+1$ (IE I)}

With the choice  
\begin{eqnarray}
M=\frac {(l_1+l_1^*)+i(m_1+m_1^*)}{m_1^*-il_1^*},\nonumber
\end{eqnarray}
the corresponding $1$-SS of IE I takes the form
\numparts
\begin{eqnarray}
 S^+ &=2E\frac {
(l_ {1R}^2m_ {1R}+m_ {1R}^2l_ {1I}+L)
 \exp {i\chi_ {1I}} \mbox {sech} \chi_ {1R}} {A+2B\tanh 
{\chi_ {1R}}+C\tanh^2{\chi_{1R}}},\\
S_3 &=1-\frac {2(l_ {1R}^2-m_ {1R}^2)^3 \mbox {sech}^2 \chi_ {1R}}{A+2B\tanh 
{\chi_ {1R}}+C\tanh^2{\chi_ {1R}}},
\end{eqnarray}
\endnumparts
where
\begin{eqnarray}
\fl E=(l_{1R}+im_{1R})(l_{1R}^2-m_{1R}^2),\nonumber \\
\nonumber \\
\fl L=i(m_{1I}m_{1R}^2-l_{1R}^3)+(m_{1R}^3+l_{1I}l_{1R}^2+i(m_{1I}l_{1R}^2-l_{1R}m_ {1R}^2))
\tanh {\chi_{1R}},\nonumber\\
\nonumber \\
\fl A=2l_{1R}^6-2m_{1R}^2(l_{1R}^4+l_{1R}^3m_{1I})+m_{1R}^4l_{1I}^2+2l_{1R}^2l_{1I}m_ {1R}^3+
m_{1I}^2m_{1R}^4+3l_{1R}^2m_{1R}^4-m_{1R}^6,\nonumber \\
\nonumber \\
\fl B=l_{1R}^2m_{1R}^2(m_{1R}^2+l_{1I}^2+m_{1I}^2+l_{1R}^2)+l_{1R}^3l_{1I}m_{1R}
+l_{1I}(m_{1R}^5-l_{1R}^5)-l_{1R} m_{1I}m_{1R}^4\nonumber \\
\nonumber \\
\fl C=2m_{1R}^6+2l_{1I}l_{1R}^2m_{1R}^3+l_{1R}^4l_{1I}^2+m_{1I}^2l_{1R}^4+
3m_{1R}^2l_{1R}^4-2l_{1R}^2m_{1R}^4-\nonumber \\
2m_{1I}l_{1R}^3m_{1R}^2-l_ {1R}^6.\nonumber
\end{eqnarray}
\underline{Case(ii): $\sigma^2=-1$ (IE II)}

Choosing 
\begin{eqnarray}
M=\frac {(l_1+l_1^*)+i(m_1+m_1^*)}{l_1^*-im_1^*},\nonumber
\end{eqnarray}
the $1$-SS of IE II takes the form
\numparts
\begin{eqnarray}
\fl S^+=-2
\frac{(l_ {1R}+im_ {1R})+(m_ {1I}-il_ {1I}+(l_ {1R}+im_ {1R})
\tanh {\chi_ {1R}}) \exp {i\chi_{1I}} \mbox{sech} \chi_ {1R}}
{l_ {1R}^2+l_ {1I}^2+
2(l_ {1R}m_ {1I}-l_ {1I}m_ {1R})\tanh{\chi_{1R}+(m_ {1R}^2+m_
 {1I}^2)\tanh^2{\chi_ {1R}}}},\\
\fl S_3=1-\frac{2(l_ {1R}^2+m_ {1R}^2) \mbox {sech}^2 \chi_ {1R}} {l_
 {1R}^2+l_ {1I}^2+
2(l_ {1R}m_ {1I}-l_ {1I}m_ {1R})\tanh {\chi_ {1R}+(m_ {1R}^2+m_
 {1I}^2)\tanh^2{\chi_ {1R}}}}.
\end{eqnarray}
\endnumparts

\subsubsection{2-soliton solution}

To construct $2$-SS, we take $N=2$ in eq.~(2.9). Then, $g_1$ takes the form
\begin{eqnarray}
g_1=\exp {\chi_1}+\exp {\chi_2}.
\end{eqnarray}
Substituting (2.13) in (2.8) and after some calculations we obtain
\numparts
\begin{eqnarray}
\fl f_2=M_ {11}\exp {(\chi_1+\chi_1^*)}+M_ {12}\exp {(\chi_2+\chi_1^*)}+M_ {21}\exp 
{(\chi_1+\chi_2^*)}+\nonumber \\
M_ {22}\exp {(\chi_2+\chi_2^*)},\\
\fl g_3=L_ {112}\exp {(\chi_1+\chi_1^*+\chi_2)}+L_ {122}\exp
{(\chi_1+\chi_2+\chi_2^*)},\\
\fl f_4=K \exp {(\chi_1+\chi_1^*+\chi_2+\chi_2^*)},
\end{eqnarray}
\endnumparts
where 
\[
M_ {rs}=\frac {\sigma^2m_s^2-l_s^2}{(l_r^*+l_s)^2-\sigma^2(m_r^*+m_s)^2},  \quad
r,s=1,2,
\]
\[
L_{rst}=\frac{(\sigma^2m_s^{*2}-l_s^{*2})((l_r-l_t)^2-\sigma^2(m_r-m_t)^2)} 
{((l_r+l_s^*)^2-\sigma^2(m_r+m_s^*)^2) ((l_t+l_s^*)^2-\sigma^2(m_t+m_s^*)^2)}
,  \quad t=1,2,\]
\[
K=\frac {((l_1^*-l_2^*)^2-\sigma^2(m_1^*-m_2^*)^2)
((l_1-l_2)^2-\sigma^2(m_1-m_2)^2)P} 
{((l_1+l_1^*)^2-\sigma^2(m_1+m_1^*)^2)
((l_2+l_1^*)^2-\sigma^2(m_2+m_1^*)^2)},\]
\[
P=\frac {\sigma^2m_1^2-l_1^2}{(l_1+l_2^*)^2-\sigma^2(m_1+m_2^*)^2}
\frac {\sigma^2m_2^2-l_2^2}{(l_2+l_2^*)^2-\sigma^2(m_2+m_2^*)^2}.
\]
Making use of the fact that now $\omega =\frac {g}{f}=\frac {\epsilon
g_1+\epsilon^3g_3}{1+\epsilon^2f_2+\epsilon^4f_4}$, and the relations (2.1) for
the spin variables in terms of $\omega$, the spin two soliton solution can be
written explicitly. 

\subsubsection{N-soliton solution}

Finally, by taking $g_1$ as in eq.~(2.9) and extending the above procedure, one
can obtain the $N$-SS as
\numparts
\begin{eqnarray}
g=\sum_ {\mu_j=0,1}^" \exp
\{\sum_ {i<j}^ {2N}\phi(i,j)\mu_i\mu_j+\sum_ {i=1}^ {2N}\mu_i[\chi_i+\psi(i)]\},
\end{eqnarray}
where
\begin{eqnarray}
\psi(i)=\left\{
\begin{array}{ll}
\log (\sigma^2m_i^2-l_i^2) & \mbox{for}\;\; i=N+1,\ldots,2N, \cr
 0 & \mbox{otherwise},
\end{array}
 \right. \nonumber \\
\sum_ {\mu_i=0,1}^" \mbox {means}
\sum_ {i=1}^N\mu_i=1+\sum_ {i=1}^N\mu_ {i+N},\nonumber
\end{eqnarray}
and 
\begin{eqnarray}
f=\sum_ {\mu_j=0,1}^{'}\exp
\{\sum_ {i<j}^ {2N}\phi (i,j)\mu_i\mu_j+\sum_ {i=1}^ {2N}\mu_i[\chi_i+
\psi'(i)]\},
\end{eqnarray}
where
\[
\chi_i=l_ix+m_iy+n_it+\chi_i(0),\]
\[n_i=i(l_i^2+\sigma^2m_i^2),\quad \chi_{i+N}=\chi_i^*, l_{i+N}=l_i^*,
m_ {i+N}=m_i^*, n_ {i+N}=n_i^*,\]
\begin{eqnarray}
\fl \phi(i,j)=\left\{ 
\begin{array}{cc}
-\log ((l_i+l_j)^2-\sigma^2(m_i+m_j)^2) & \mbox{for}\; i=1, \ldots ,N\;\; 
 \mbox{and}\; j=N+1,\ldots,2N, \\
 \log ((l_i-l_j)^2-\sigma^2(m_i-m_j)^2) & \mbox{for}\; i=1, \ldots ,N\;\;
 \mbox{and}\;j=1,\ldots,N, 
\end{array}
\right. \nonumber
\end{eqnarray}
\begin{eqnarray}
\psi'(i)=\left\{ 
\begin{array}{cc}
 \log (\sigma^2m_i^2-l_i^2) & \mbox{for} \; i=1,\ldots,N, \\
 0 & \mbox{otherwise},
\end{array} 
\right. \nonumber
 \end{eqnarray}
 \endnumparts
\[ \sum_ {\mu_i=0,1}^{'}\mbox {means} 
\sum_ {i=1}^N\mu_i=\sum_ {i=1}^N\mu_ {i+N}.
\]

\section{Bilinearization and solution of IE I in terms of light cone
coordinates: The role of boundaries}

In the case of most of the (2+1) dimensional integrable nonlinear evolution
equations like Davey-Stewartson I, modified Kadomtsev-Petviashvili,
and Nizhnik-Veselov-Novikov equations\cite{hie,RR97a,RR97b,RR97c},  the
bilinearized forms are first
transformed into systems of linear partial differential equations(pdes)
while using the Hirota's direct method. By solving these linear pdes, one can construct
the line solitons(as we have done in section 2) which are localized everywhere
except along particular lines. Looking at the nature of line solitons, the
presence of two nonparallel ghost solitons\cite{hie}  (solitons which are
visible only in the absence of the physical field) are identified. As a dromion
is the two soliton solution made out of two nonparallel ghost solitons, they
can be embedded in the two soliton solution to generate a (1,1) dromion. It
implies that the solution consists of one bound state each in the $x$ and $y$
directions. This can be directly extended to generate multidromions. If we
follow the same procedure for IE, we see from eq.~(2.11) or eq.~(2.12) that the
two nonparallel ghost solitons are absent here. Hence we cannot construct
dromions from the bilinearized form (2.6). 

In the following, we analyse the IE I in a different frame of reference
consisting of light cone coordinates $\xi$ and $\eta$, which are defined as
\begin{eqnarray}
\xi=\frac {1}{2} (y+x),\quad \eta =\frac {1}{2} (y-x). 
\end{eqnarray}
Correspondingly, eq.~(1.1) takes the form (after rescaling $-\frac
{t}{2}\longrightarrow t' ; t'\longrightarrow t$)
\numparts
\begin{eqnarray}
\fl \vec S_t(\xi,\eta,t)=\vec S\wedge (\vec S_ {\xi\xi}+\vec S_ {\eta
\eta})+\phi_ {\xi}\vec S_ {\xi}-\phi_ {\eta} \vec S_ {\eta},\\
     \phi_ {\xi \eta} = \vec S\cdot \vec S_ {\xi} \wedge \vec S_ {\eta}.
\end{eqnarray}
\endnumparts
The form of eq.~(3.2b) suggests that one can redefine the scalar field variable $\phi$ as
\begin{eqnarray}
\phi(\xi,\eta,t)=\Phi(\xi,\eta,t)+\int m_1(\xi,t) d\xi+\int m_2(\eta,t) 
d\eta, 
\end{eqnarray}
where $m_1(\xi,t)$ and $m_2(\eta,t)$ are the boundaries (arbitrary functions in
the indicated variables). Eq.~(3.2) can therefore be rewritten as 
\numparts
\begin{eqnarray}
\fl \vec S_t(\xi,\eta,t)=\vec S\wedge (\vec S_ {\xi\xi}+\vec S_ {\eta
\eta})+(\Phi_ {\xi}+m_1(\xi,t))\vec S_ {\xi}-(\Phi_ {\eta}+m_2(\eta,t)) \vec S_ {\eta},\\
\Phi_ {\xi \eta}= \vec S\cdot \vec S_ {\xi} \wedge \vec S_ {\eta}.
\end{eqnarray}
\endnumparts
In terms of the stereographic variable $\omega$ given by eq.~(2.1), eq.~(3.4)
takes the form
\numparts
\begin{eqnarray}
\fl i\omega_t+\omega_ {\xi \xi}+\omega_ {\eta \eta}
-\frac{2\omega^*(\omega_ {\xi}^2+\omega_ {\eta}^2)}{1+|\omega|^2}
-i(\Phi_ {\xi}+m_1(\xi,t))
\omega_ {\xi}-i(\Phi_ {\eta}+m_2(\eta,t))\omega_ {\eta}=0,\label{}\\
\fl \Phi_ {\xi \eta}=\frac {4i}{(1+|\omega|^2)^2}(\omega_ {\xi}^*\omega_
 {\eta}-\omega_ {\xi}\omega_ {\eta}^*).
\end{eqnarray}
\endnumparts
After introducing the transformation $\omega =\frac {g}{f}$, the bilinear
representations of eq.~(3.4) can be written as
\numparts
\begin{eqnarray}
(iD_t-D_{\xi}^2-D_{\eta}^2-im_1(\xi,t)D_{\xi}+im_2(\eta,t)D_{\eta})(f^*\cdot g)=0,\\
(iD_t-D_{\xi}^2-D_{\eta}^2-im_1(\xi,t)D_{\xi}+im_2(\eta,t)D_{\eta})(f^*\cdot f-g^*\cdot g)=0,\\
\fl D_ {\xi}(D_ {\eta}(f^*\cdot f+g^*\cdot g))\cdot (f^*f+g^*g)=
D_{\eta}(D_{\xi}(f^*\cdot f+g^*\cdot g))\cdot (f^*f+g^*g),
\end{eqnarray}
so that 
\begin{eqnarray}
\Phi_ {\xi}=-2i \frac {D_ {\xi}(f^*\cdot f+g^*\cdot g)} {(f^*f+g^*g)},\\
\Phi_ {\eta}=-2i \frac {D_ {\eta}(f^*\cdot f+g^*\cdot g)} {(f^*f+g^*g)}.
\end{eqnarray}
\endnumparts
One observes now the explicit introduction of the boundaries $m_1(\xi,t)$ and
$m_2(\eta,t)$  into the bilinearized form (3.6), which turns out to be crucial
to obtain localized solutions. Also eq.~(3.6c) is a consequence of the
compatibility between eqs.~(3.6d) and (3.6e). It may also be noted that a
similar introduction of arbitrary functions in the bilinear form was necessiated
for the (2+1) dimensional sine-Gordan equation in order to obtain localized
solutions\cite{RR97c}.

Expanding now the functions $g$ and $f$ as  power series, 
\numparts
\begin{eqnarray}
g=\epsilon g_1+ \epsilon^3 g_3+\epsilon^5 g_5+\ldots, \\
f=1+\epsilon^2 f_2+\epsilon^4 f_4+\epsilon^6 f_6+\ldots  
\end{eqnarray}
\endnumparts
and substituting them into eqs.~(3.6a)-(3.6c), we will obtain the 
following set of linear pdes by equating the various powers of
$\epsilon$:
\numparts
\begin{eqnarray}
\fl \epsilon:
(iD_t-D_ {\xi}^2-D_ {\eta}^2-im_1(\xi,t)D_ {\xi}+im_2(\eta,t)D_ {\eta})(1\cdot
g_1)=0,\label {eq1}\\
\fl \epsilon^2:
(iD_t-D_ {\xi}^2-D_ {\eta}^2-im_1(\xi,t)D_ {\xi}+im_2(\eta,t)D_ {\eta})(1 \cdot
f_2+f_2^*\cdot 1-g_1^*\cdot g_1)=0,\\
\fl \quad D_ {\xi}(D_ {\eta}(1 \cdot
f_2+f_2^*\cdot 1-g_1^*\cdot g_1)\cdot 1)+D_ {\eta}(D_ {\xi}(1 \cdot
f_2+f_2^*\cdot 1-g_1^*\cdot g_1) \cdot 1=0
\end{eqnarray}
\endnumparts
and so on. Let us solve the above linear equations to obtain the solutions to
IE I. 

Eq.~(3.8a) can be rewritten as
\begin{eqnarray}
ig_{1t}+g_ {1\xi\xi}+g_ {1\eta \eta}-im_1(\xi,t)g_ {1\xi}+im_2(\eta,t)g_ {1\eta}=0.
\end{eqnarray} 
Let us try a variable separation, by postulating 
\begin{eqnarray}
g_1=p(\xi,t) q(\eta,t), 
\end{eqnarray}
where $p$ and $q$ are complex functions of the indicated arguments. Now,
eq.~(3.9) becomes
\begin{eqnarray}
q(ip_t+p_ {\xi\xi}-im_1(\xi,t)p_ {\xi})+p(iq_t+q_ {\eta \eta}+im_2(\eta,t)q_ {\eta})=0.
\end{eqnarray}
The above equation suggests that we should have
\numparts
\begin{eqnarray}
ip_t+p_ {\xi\xi}-im_1(\xi,t)p_ {\xi}=k_1p,\\
iq_t+q_ {\eta \eta}+im_2(\eta,t)q_ {\eta}=-k_1q,
\end{eqnarray}
\endnumparts
where $k_1$ is a constant. Redefining $p=\hat p\exp {-ik_1t}$ and 
$q=\hat q\exp{ik_1t}$ and dropping the hats, eq.~(3.12) becomes
\numparts
\begin{eqnarray}
ip_t+p_ {\xi\xi}-im_1(\xi,t)p_ {\xi}=0,\\
iq_t+q_ {\eta \eta}+im_2(\eta,t)q_ {\eta}=0.
\end{eqnarray}
\endnumparts
Simplifying now eq.~(3.8c), we will get
\begin{eqnarray}
f^*_{2\xi \eta}-f_{2\xi \eta}=g_1^*g_{1\xi \eta}-g_1g^*_{1\xi \eta}.
\end{eqnarray}
Let us write $f_2=f_{2R}+if_{2I}$, where $f_{2R}$ and $f_{2I}$ are the real and
imaginary parts of the function $f_2$, respectively. Eq.~(3.14) now becomes 
\begin{eqnarray}
f_{2I\xi \eta}=\frac {1}{2i} (g_1g^*_{1\xi \eta}-g_1^*g_{1\xi \eta}).\nonumber 
\end{eqnarray}
On carrying out integrations with respect to $\xi$ and $\eta$, we get  
\begin{eqnarray}
f_{2I}=\frac {1}{2i}
\int\int (g_1g_{1\xi \eta}^*-g_1^*g_{1\xi \eta}) d\xi d\eta+h(\xi,t)+r(\eta,t),
\end{eqnarray}
where $h(\xi,t)$ and $r(\eta,t)$ are arbitrary functions in the indicated
arguments. Substitution of $g_1$ and $f_2$ in eq.(3.8b) gives rise to
\begin{eqnarray}
\fl \nabla^2f_{2R}\equiv f_{2R\xi\xi}+f_{2R\eta\eta}=f_ {2It}-m_1(\xi,t)f_ {2I\xi}
+m_2(\eta,t)f_ {2I\eta}-(p_ {\xi}p_ {\xi}^*qq^*+pp^*q_ {\eta}q_ {\eta}^*),
\end{eqnarray}
which is nothing but the Poisson's equation in two dimensions for $f_{2R}$. For
the given boundaries $m_1(\xi,t)$ and $m_2(\eta,t)$, once the eqs.~(3.13a) and
(3.13b) are solved for $p$ and $q$ respectively, the right hand sides of eqs.~(3.15) and (3.16) can be found out. Hence they can be solved to obtain $f_{2I}$
and $f_{2R}$. At this point, we may point out that exactly the same kind of
linear equations (3.12a) and (3.12b) have appeared both in the case of IST
method\cite{BGK:1992} and binary Darboux transformation method\cite{IMAI}. Now these equations have
appeared from a different perspective, namely from the point of view of
bilinearization method. 

Now the general solution for $\omega$ can be written as
\begin{eqnarray}
\omega=\frac{g}{f}=\frac {p(\xi,t) q(\eta,t)}{1+f_{2R}+if_{2I}},
\end{eqnarray}
where the functions $p$, $q$, $f_{2I}$ and $f_{2R}$ are to be determined from 
equations (3.13a), (3.13b), (3.15) and (3.16), respectively. Here the main task lies in
solving eq.~(3.13) which is also associated with the linear problem of the following 
modified Kadomtsev-Petviashvili equation(mKP)\cite{BGKstu}: 
\begin{eqnarray}
u_t+u_{xxx}+6u^2 u_x-12 \partial_x^{-1}u_{yy}+12u_x\partial_x^{-1}u_y=0.
\end{eqnarray}
The mKP eq.~(3.15) is equivalent to the compatibility condition for the linear
system,  
\numparts
\begin{eqnarray}
(2i \partial_y+\partial_x^2+2iu\partial_x) \psi=0,\\
(i\partial_t+4 \partial_x^3+12 i u \partial_x^2+(6 i u_x+12 \partial_x^{-1}
u_y-6 u^2)\partial_x+a)\psi=0,
\end{eqnarray}
\endnumparts
where $a$ is an arbitrary constant. By comparing eq.~(3.13) and eq.~(3.19), it is
seen that the proper solutions of eq.~(3.13) are obtained by dropping the time
dependence in the various types of solutions such as line solitons, line lumps
and line breathers of eq.~(3.19) and changing $y\rightarrow t$. Since we have
two independent problems in eq.~(3.13), large number of classes of exact solutions of IE
I, which include rationally  localized, rationally-exponentially
localized  and fully exponentially  localized structures, are possible. We
present the details of some of them in the next section.

\section{Localized coherent structures of IE I: Time dependent boundaries}

\subsection{Lump-lump boundaries: Algebraically decaying structures}

In this section, we choose specific forms of the boundaries $m_1(\xi,t)$ and
$m_2(\eta,t)$ in eq.~(3.8) to obtain localized solutions. For the algebraically
decaying boundaries of the form (for easy comparison, we follow the notations of
ref. \cite{BGK:1992})
\numparts
\begin{eqnarray}
m_1(\xi,t)=-\frac {2\alpha}{(\xi-\frac {2t}{\alpha}+c_1)^2+\frac {\alpha^2}{4}},\label {lb1}\\
m_2(\eta,t)=\frac {2\beta}{(\eta-\frac {2t}{\beta}+c_2)^2+\frac {\beta^2}{4}},\label {lb2}
\end{eqnarray}
\endnumparts
where $\alpha$, $\beta$, $c_1$ and $c_2$ are real constants, the solutions to
eqs.~(3.13) are obtained as 
\numparts
\begin{eqnarray}
p(\xi,t)=\frac {\exp {i(\frac {\xi}{\alpha}-\frac {t}{\alpha^2})}}{\xi-\frac
{2t}{\alpha}+c_1-\frac {i\alpha}{2}},\\
q(\eta,t)=\frac {\exp {i(\frac {\xi}{\beta}-\frac {t}{\beta^2})}}{\eta-\frac
{2t}{\beta}+c_2-\frac {i\beta}{2}}.
\end{eqnarray}
\endnumparts
Substitution of eq.~(4.2) in (3.15) leads to
\begin{eqnarray}
f_{2I}=\frac {1}{2\alpha \beta} \frac {\beta (\xi-\frac {2t}{\alpha}+c_1)+\alpha 
(\eta-\frac
{2t}{\beta}+c_2)} {((\xi-\frac {2t}{\alpha}+c_1)^2+\frac {\alpha^2}{4})
((\eta-\frac {2t}{\beta}+c_2)^2+\frac {\beta^2}{4})},
\end{eqnarray}
wherein $h(\xi,t)=0$ and $r(\eta,t)=0$ and substituting the values of $p$, $q$ and $f_{2I}$ 
in eq.~(3.16) and solving the
resultant Poisson's equation for $f_{2R}$, we get 
\begin{eqnarray}
f_{2R}=\frac {1}{\alpha \beta} \frac {(\xi-\frac {2t}{\alpha}+c_1)+(\eta-\frac
{2t}{\beta}+c_2)-\frac {\alpha \beta}{4}} {((\xi-\frac {2t}{\alpha}+c_1)^2+\frac {\alpha^2}{4})
((\eta-\frac {2t}{\beta}+c_2)^2+\frac {\beta^2}{4})}.
\end{eqnarray}
Therefore,
\begin{eqnarray}
f_2=f_{2R}+if_{2I}=\frac {1}{\alpha \beta (\xi-\frac
{2t}{\alpha}+c_1-\frac {i\alpha}{2}) (\eta-\frac
{2t}{\beta}+c_2-\frac {i\beta}{2})}.
\end{eqnarray}
Hence, from eq.~(3.17), we have 
\begin{eqnarray}
\omega=\frac {\exp {i(\frac {\xi}{\alpha}-\frac {t}{\alpha^2}+\frac 
{\xi}{\beta}-\frac {t}{\beta^2})}} {\frac {1} {\alpha \beta}+(\xi-\frac 
{2t}{\alpha}+c_1-\frac {i\alpha}{2}) (\eta-\frac {2t}{\beta}+c_2-\frac 
{i\beta}{2})}.
\end{eqnarray}
Using eq.~(2.1), the spin components corresponding to this form of $\omega$ are given below:
\numparts
\begin{eqnarray}
\fl S^+=2\frac {({\frac {1} {\alpha \beta}+(\xi-\frac 
{2t}{\alpha}+c_1-\frac {i\alpha}{2}) (\eta-\frac {2t}{\beta}+c_2-\frac 
{i\beta}{2})}) {\exp {i(\frac {\xi}{\alpha}-\frac {t}{\alpha^2}+\frac 
{\xi}{\beta}-\frac {t}{\beta^2})}}} {A^2+B^2},\\
S_3=1-\frac {2}{A^2+B^2},
\end{eqnarray}
\endnumparts
where
\[
A=\frac {1} {\alpha \beta}+(\xi-\frac 
{2t}{\alpha}+c_1) (\eta-\frac {2t}{\beta}+c_2)+\frac 
{\alpha \beta}{4}
\] 
and \[
B=\frac {\alpha}{2} (\eta-\frac {2t}{\beta}+c_2)-\frac {\beta}{2} (\xi-\frac 
{2t}{\alpha}+c_1).\] 
The solution (4.7) decays as $\frac {1}{\xi \eta}$ as
$\xi^2+\eta^2\longrightarrow \infty$ and moves with the velocity $V=(V_{\xi},
V_{\eta})=(\frac {2}{\alpha}, \frac {2}{\beta})$. A snap shot of $S_3$ is
plotted in fig.~(1)

\begin{figure}[h]
\resizebox{.8\textwidth}{!}{
\includegraphics{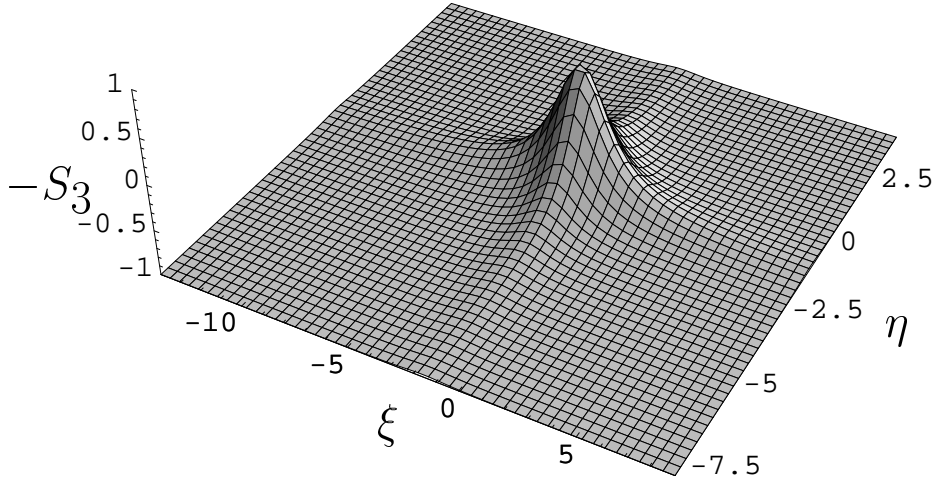}}
\caption{Algebraically decaying lump structure}
\end{figure}

\subsection{Soliton-soliton boundaries: Exponentially decaying dromions}

Let us now choose the boundaries $m_1(\xi,t)$ and $m_2(\eta,t)$ as the following 
line solitons(again using the notations of \cite{BGK:1992}):
\numparts
\begin{eqnarray}
m_1(\xi,t)=\frac {\frac {8\mu_I}{|\mu|^2} \exp {\frac {2\mu_I}{|\mu|^2} \hat
\xi}} {(1+\frac {\mu_R}{\mu_I} \exp {\frac {2\mu_I}{|\mu|^2} \hat \xi})^2+\exp 
{\frac {4\mu_I}{|\mu|^2} \hat \xi}},\;\;\mbox{($\mu=\mu_R+i\mu_I$)}\\
m_2(\eta,t)=-\frac {\frac {8\lambda_I}{|\lambda|^2} \exp {\frac {2\lambda_I}{|\lambda|^2} \hat
\eta}} {(1+\frac {\lambda_R}{\lambda_I} \exp {\frac {2\lambda_I}{|\lambda|^2}
\hat \eta})^2+\exp {\frac {4\lambda_I}{|\lambda|^2} \hat \eta}},\mbox{(
$\lambda=\lambda_R+i\lambda_I$)}
\end{eqnarray}
\endnumparts
where $\hat \xi=\xi-\frac {2\mu_R}{|\mu|^2}t+\xi_{0}, \hat \eta=\eta-\frac
{2\lambda_R}{|\lambda|^2}t+\eta_{0}$ and $\mu_R$, $\mu_I$, $\lambda_R$,
$\lambda_I$, $\xi_{0}$ and $\eta_{0}$ are real constants. The suffices $R$ and $I$ are used to denote the
real and imaginary parts respectively. On solving eqs.~(3.13), (3.15) and (3.16),
we get the following expressions for $p(\xi,t)$, $q(\eta,t)$ and $f(\xi,
\eta,t)$:
\numparts
\begin{eqnarray}
p(\xi,t)=\frac {\exp {i(\frac {\xi}{\mu} -\frac {t}{\mu^2})}} {(1+\frac 
{\mu^*}{\mu_I} \exp {\frac {2\mu_I}{|\mu|^2} \hat \xi})},\\
q(\eta,t)=\frac {\exp {i(\frac {\eta}{\lambda}-\frac {t}{\lambda^2})}}
{(1+\frac {\lambda^*}{\lambda_I} \exp {\frac {2\lambda_I}{|\lambda|^2}
\hat \eta})},\\
f_{2I}=\frac {1}{4} \frac {(1+\frac {\lambda_R}{\lambda_I} \exp {\frac {2\lambda_I}{|\lambda|^2}
\hat \eta}) \exp {\frac {2\mu_I}{|\mu|^2} \hat \xi} + (1+\frac 
{\mu_R}{\mu_I} \exp {\frac {2\mu_I}{|\mu|^2} \hat \xi}) \exp {\frac {2\lambda_I}{|\lambda|^2} \hat \eta} } {|1+\frac {\mu}{\mu_I}
\exp {\frac {2\mu_I}{|\mu|^2} \hat \xi}|^2 \;\;\;|1+\frac {\lambda}{\lambda_I} \exp {\frac {2\lambda_I}{|\lambda|^2}
\hat \eta}|^2},\\
f_{2R}=\frac {(1+\frac {\mu_R}{\mu_I} \exp {\frac {2\mu_I}{|\mu|^2} \hat \xi})
(1+\frac {\mu_R}{\mu_I} \exp {\frac {2\mu_I}{|\mu|^2} \hat \xi})-\exp {\frac {2\mu_I}{|\mu|^2} \hat \xi+\frac {2\lambda_I}{|\lambda|^2} \hat
\eta}} {4 |1+\frac {\mu}{\mu_I}
\exp {\frac {2\mu_I}{|\mu|^2} \hat \xi}|^2\;\;\; |1+\frac {\lambda}{\lambda_I} \exp {\frac {2\lambda_I}{|\lambda|^2}
\hat \eta}|^2}.
\end{eqnarray}
\endnumparts
Consequently, we have 
\begin{eqnarray}
\omega=\frac {\exp {i(\frac {\xi}{\mu} -
\frac {t}{\mu^2}+\frac {\eta}{\lambda}-\frac {t}{\lambda^2})}}{(1+\frac 
{\mu^*}{\mu_I} \exp {\frac {2\mu_I}{|\mu|^2} \hat \xi}) (1+\frac {\lambda^*}{\lambda_I} \exp {\frac {2\lambda_I}{|\lambda|^2}
\hat \eta})+\frac {1}{4}} .
\end{eqnarray} 
The spin components corresponding to this form of $\omega$ are given below:
\numparts
\begin{eqnarray}
\fl S^+=\frac {2{((1+\frac 
{\mu^*}{\mu_I} \exp {\frac {2\mu_I}{|\mu|^2} \hat \xi}) (1+\frac {\lambda^*}{\lambda_I} \exp {\frac {2\lambda_I}{|\lambda|^2}
\hat \eta})+\frac {1}{4})} \exp {i(\frac {\hat \xi}{\mu} -
\frac {t}{\mu^2}+\frac {\hat \eta}{\lambda}-\frac {t}{\lambda^2})}}
{C^2+D^2+\exp ({\frac {2\mu_I}{|\mu|^2} \hat \xi+\frac {2\lambda_I}
{|\lambda|^2} \hat \eta})},\\
\fl S_3=1-\frac {2\exp ({\frac {2\mu_I}{|\mu|^2} \hat \xi+\frac {2\lambda_I}
{|\lambda|^2} \hat \eta})} {C^2+D^2+\exp ({\frac {2\mu_I}{|\mu|^2} \hat \xi+\frac {2\lambda_I}
{|\lambda|^2} \hat \eta})},
\end{eqnarray}
\endnumparts
where
\[ \fl 
C=1+\frac{1}{4}+\frac{\mu_R}{\mu_I} \exp {\frac{2\mu_I}{|\mu|^2} \hat \xi}+
\frac {\lambda_R}{\lambda_I} \exp {\frac{2\lambda_I}{|\lambda|^2} \hat
\eta}+\frac {1}{\lambda_I\mu_I} (\lambda_R\mu_R-\lambda_I\mu_I)
 \exp ({\frac {2\mu_I}{|\mu|^2} \hat \xi+\frac {2\lambda_I}{|\lambda|^2} 
 \hat \eta}),
\]
\[ \fl 
D=\exp {\frac {2\mu_I}{|\mu|^2} \hat \xi}+\exp {\frac {2\lambda_I}
{|\lambda|^2} \hat \eta}+(\frac {\lambda_R} {\lambda_I}+\frac {\mu_R}{\mu_I})
\exp ({\frac {2\mu_I}{|\mu|^2} \hat \xi+\frac {2\lambda_I}
{|\lambda|^2} \hat \eta}).
\]
The solution (4.11) decays exponentially (see fig.~(2)) in all directions on the plane 
$\xi$, $\eta$ and moves with the velocity $V=(V_{\xi},
V_{\eta})=(\frac {2\mu_R}{|\mu|^2}, \frac {2\lambda_R}{|\lambda|^2})$.

\begin{figure}[h]
\resizebox{.8\textwidth}{!}{
\includegraphics{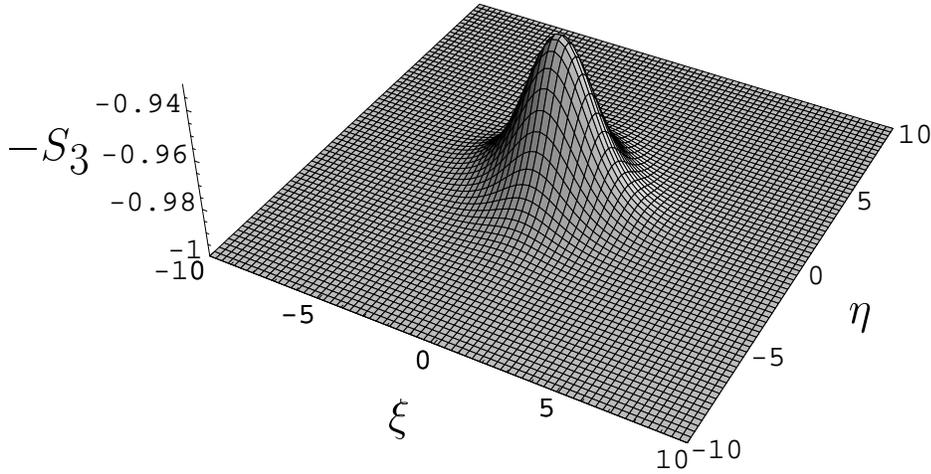}}
\caption{A dromion}
\end{figure}

\subsection{Lump-line soliton boundaries: Rationally-exponentially decaying
nature}

Taking the choice of the rational lump as the boundary $m_1(\xi,t)$ and of the
line soliton as the boundary $m_2(\eta,t)$, that is
\numparts
\begin{eqnarray}
m_1(\xi,t)=-\frac {2\alpha}{(\xi-\frac {2t}{\alpha}+c_1)^2+\frac
{\alpha^2}{4}},\label {lb11}\\
m_2(\eta,t)=-\frac {\frac {8\lambda_I}{|\lambda|^2} \exp {\frac {2\lambda_I}{|\lambda|^2} \hat
\eta}} {(1+\frac {\lambda_R}{\lambda_I} \exp {\frac {2\lambda_I}{|\lambda|^2}
\hat \eta})^2+\exp {\frac {4\lambda_I}{|\lambda|^2} \hat \eta}},
\end{eqnarray}
\endnumparts
and proceeding as before, we can find the solution of $\omega$ as
\begin{eqnarray}
\omega=\frac {\exp {i(\frac {\xi}{\alpha}-\frac {t}{\alpha^2}+\frac {\eta}
{\lambda}-\frac {t}{\lambda^2})}} {\frac {1}{2\alpha}+(\xi-\frac {2t}{\alpha}+c_1-\frac {i\alpha}{2})
(1+\frac {\lambda^*}{\lambda_I} \exp {\frac {2\lambda_I}{|\lambda|^2}\hat \eta})
}.
\end{eqnarray}
The spin components in this case are
\numparts
\begin{eqnarray}
\fl S^{+}=\frac {2({\frac {1}{2\alpha}+(\xi-\frac {2t}{\alpha}+c_1+\frac {i\alpha}{2})+
(1+\frac {\lambda}{\lambda_I} \exp {\frac {2\lambda_I}{|\lambda|^2}\hat \eta}))
} \exp {i(\frac {\xi}{\alpha}-\frac {t}{\alpha^2}+\frac {\hat \eta}
{\lambda}-\frac {t}{\lambda^2})}} {E^2+F^2+\exp {\frac
{2\lambda_I}{|\lambda|^2}\hat \eta}},\\
\fl S_3=1-\frac {2\exp {\frac {2\lambda_I}{|\lambda|^2}\hat \eta}} {E^2+F^2+\exp {\frac
{2\lambda_I}{|\lambda|^2}\hat \eta}},
\end{eqnarray}
\endnumparts
where
\[
E=(\xi-\frac {2t}{\alpha}+c_1)-\frac {\alpha}{2} 
\exp {\frac {2\lambda_I}{|\lambda|^2}\hat \eta}+\frac {\lambda_R}{\lambda_I} 
 (\xi-\frac {2t}{\alpha}+c_1) \exp {\frac {2\lambda_I}{|\lambda|^2}\hat \eta}  +\frac {1}{2\alpha}
\]
\[
F=\frac {\alpha}{2}(1+\frac {\lambda_R}{\lambda_I} \exp {\frac {2\lambda_I}{|\lambda|^2}
\hat \eta})+(\xi-\frac {2t}{\alpha}+c_1)\exp {\frac {2\lambda_I}{|\lambda|^2}
\hat \eta}.\]
The above solution (4.14) decays rationally in the $\xi$ direction and 
exponentially in the $\eta$ direction on the plane $\xi$, $\eta$ and it moves
with the velocity $V=(V_{\xi},V_{\eta})=(\frac {2}{\alpha}, \frac {2\lambda_R}
{|\lambda|^2})$ (see fig.~(3)).

\begin{figure}[h]
\resizebox{.8\textwidth}{!}{
\includegraphics{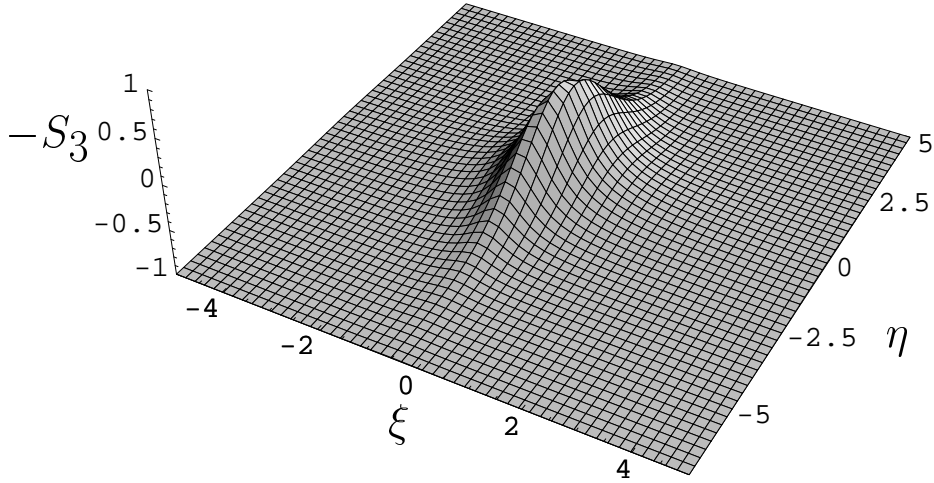}}
\caption{Rationally-exponentially deacying structure}
\end{figure}

We can take the other choice also, that is $m_1(\xi,t)$ as the plane soliton
boundary and $m_2(\eta,t)$ as the rational boundary. Replacing $\xi$ by $\eta$,
$\alpha$ by $\beta$, $\hat \eta$ by $\hat \xi$ and $\lambda$ by $\mu$ in (4.12)
and (4.13), we will get the solution corresponding to this case.

\section{Stationary boundaries}

In the case of stationary boundaries, that is $m_1(\xi,t)=m_1(\xi)$ and
$m_2(\eta,t)=m_2(\eta)$, the function $g_1$ can be written as
\begin{eqnarray}
g_1=p'(\xi) q'(\eta) T(t)
\end{eqnarray}
where $p'$, $q'$ and $T$ are complex functions of their indicated arguments.
Substitution of this expression for $g_1$ in eq.~(3.9) leads to
\begin{eqnarray}
ip'q'T_t+q'T(p'_ {\xi\xi}-im_1(\xi)p_ {\xi}')+p'T(q_ {\eta\eta}'+im_2(\eta)q_ {\eta}')=0.
\end{eqnarray}
Eq. (5.2) suggests that
\numparts
\begin{eqnarray}
p'_ {\xi\xi}-im_1(\xi)p_ {\xi}'=-\lambda^2p'\\
q_ {\eta\eta}'+im_2(\eta)q_ {\eta}')=-\lambda^{'2}q^{'}\\
T_t+i(\lambda^2+\lambda^{'2})T=0,
\end{eqnarray}
\endnumparts
where $\lambda$ and $\lambda'$ are some complex arbitrary parameters. Eq.~(5.3c) can be
solved immediately and it gives
\begin{eqnarray}
T=\exp {-i(\lambda^2+\lambda^{'2})}t.
\end{eqnarray}
Rescaling $p'(\xi)=\hat p(\xi)\exp {\frac{i}{2}\int m_1(\xi) d\xi}$ and
$q'(\eta)=\hat q(\eta)\exp {\frac{-i}{2}\int m_2(\eta) d\eta}$ and then 
dropping the hats, eqs.~(5.3a) and (5.3b) take the form
\numparts
\begin{eqnarray}
p'_{\xi\xi}+(\lambda^2+\frac{m_1^2}{4}+i\frac{m_{1\xi}}{2})p'=0,\\
q'_{\eta \eta}+(\lambda'^2+\frac{m_2^2}{4}-i\frac{m_{2\eta}}{2})q'=0.
\end{eqnarray}
\endnumparts
The forms of $f_{2I}$ and $f_{2R}$ remain the same as in eqs.~(3.15) and (3.16),
respectively, except that here we have to replace $p$ by $p'$ and $q$ by $q'$ 
and the solution to $\omega$ is also given by (3.17). Again the above linear
equations are exactly the same as the ones obtained in ref.~\cite{BGK:1992}
through IST formalism, which are however now obtained from the bilinear form of
IE I. 

It is seen that the problems (5.3a) and (5.3b) are gauge equivalent to the
spectral problems of the Schr\"odinger type
\begin{eqnarray}
\phi_ {xx}+(\lambda^2+u^2(x)\pm iu_ {x}(x))\phi (x, \lambda)=0
\end{eqnarray}
with the very special potential\cite{BGKstu}
\begin{eqnarray}
V(x)=-(u^2(x)\pm iu_ {x}(x)),
\end{eqnarray}
where $u(x)$ is a real valued function. In this case also, the class of
coherent structures admitted by (5.6) and hence by IE I are very rich.
  
\subsection{Constant boundaries:Lump solution}

If the boundaries are constant, that is $m_1(\xi,t)=m_1$ and $m_2(\eta,t)=m_2$,
then going through the various steps indicated above, the function $\omega$ is
found out to be
\begin{eqnarray}
\omega=-\frac {\exp {i((m_1+\frac {1}{\alpha})(\xi-\frac 
{t}{\alpha})+(-m_2+\frac {1}{\beta})(\eta-\frac {t}{\beta}))}}{1-\left (\frac 
{1}{(m_1+\frac {1}{\alpha})^2}+\frac {1}{(m_2-\frac {1}{\beta})^2}\right )\frac
 {\xi^2+\eta^2}{2}}.
 \end{eqnarray}
The corresponding spin components are
 \numparts
 \begin{eqnarray}
 \fl S^+=\frac {(-2+\frac {\xi^2+\eta^2}{(m_1+\frac {1}{\alpha})^2}+\frac 
 {\xi^2+\eta^2}{(m_2-\frac {1}{\beta})^2}) \exp {i((m_1+\frac {1}{\alpha})(\xi-\frac 
{t}{\alpha})+(-m_2+\frac {1}{\beta})(\eta-\frac {t}{\beta}))}} {(1-\left (\frac 
{1}{(m_1+\frac {1}{\alpha})^2}+\frac {1}{(m_2-\frac {1}{\beta})^2}\right )\frac
 {\xi^2+\eta^2}{2})^2+1},\\
 S_3=1-\frac {2}{(1-\left (\frac 
{1}{(m_1+\frac {1}{\alpha})^2}+\frac {1}{(m_2-\frac {1}{\beta})^2}\right )\frac
 {\xi^2+\eta^2}{2})^2+1}.
 \end{eqnarray}
 \endnumparts
 In this case, the solution decays algebraically as $(\frac
 {\xi^2+\eta^2}{2})^{-1}$ as ${\xi^2+\eta^2}\longrightarrow \infty$.
 \subsection{Boundaries are absent:Line solitons}
 
 If $m_1(\xi,t)=m_2(\eta,t)=0$, then on solving equations (3.13), (3.15)-(3.17), we will
 get the following results:
 \numparts
 \begin{eqnarray}
 p=A\exp {(l\xi+il^2t)},\\
 q=B\exp {(m\eta+im^2t)},\\
 \fl f_{2I}=k_1\exp {2(l_R\xi+m_R\eta-2(l_Rl_I+m_Rm_I))};\;k_1=-AA^*BB^*\left (\frac
 {l_Rm_I+l_Im_R}{4l_Rm_R}\right ),\\
 \fl f_{2R}=k_2\exp {2(l_R\xi+m_R\eta-2(l_Rl_I+m_Rm_I))};\;k_2=AA^*BB^*\left (\frac
 {l_Im_I-l_Rm_R}{4l_Rm_R}\right ),\\
 \fl \omega=\frac {AB\exp {(l\xi+m\eta+i(l^2+m^2)t)}}{1+k_3
 \exp {2(l_R\xi+m_R\eta-2(l_Rl_I+m_Rm_I))}};\;k_3=-\frac {lm}{4l_Rm_R}AA^*BB^*,
 \end{eqnarray}
 \endnumparts
 which is nothing but the line soliton solution similar to the line solitons
 obtained in section 2 except for the difference in constant values. Here $A$
 and $B$ are complex constants. So, if the boundaries are absent to start with
 in the IE I,  eq.~(3.2) or its bilinearized version (3.6),
 we obtain line soliton solutions only. From this, one can appreciate the
 important role of boundaries to generate localized coherent structures in the
 bilinearized version of IE I.
 
 \section{Conclusions and Discussions}
 
If we look at the eqs.~(3.13), (3.15), (3.16), it is seen that we are having four
 equations for six unknowns viz., $p(\xi,t), q(\eta,t), m_1(\xi,t), m_2(\eta,t),
 f_{2R}(\xi,\eta,t)$ and $f_{2I}(\xi,\eta,t)$. Hence any two of them can be
 treated as arbitrary functions. In this case, one can manipulate these arbitrary 
 functions to get a large number of
 solutions similar to the solutions generated by Lou and his coworkers for
 equations such as Davey-Stewartson model, Nizhnik-Novikov-Veselov system and 
 dispersive long wave equation\cite{Lou}. For example, if we choose
 $p(\xi,t)=\exp {\chi_1}$ and $q(\eta,t)=\exp {\chi_2}$, where
 $\chi_1=lx+\omega_1t$ and $\chi_2=my+\omega_2t$, we will get the line
 solitons. It should be noted that we can interpret our analysis on localized
 coherent structures done in section 4 entirely in a different way. That is,
 choose the values of $p(\xi,t)$, and  $q(\eta,t)$ given in eqs. (4.2a) and
 (4.2b), then solve the eqs.~(3.13a), (3.13b), (3.15) and (3.16) to obtain the
 values of $m_1(\xi,t), m_2(\eta,t), f_{2I}(\xi,\eta,t)$ and
 $f_{2R}(\xi,\eta,t)$ respectively. Once again we arrive at the same
 algebraically decaying structure of IE I given in eq.~(4.7). Similar
 interpretation holds good for generating localized structures of other types 
 also.
 
 On the otherhand, if we take the more general solution of mKP I equation and by 
 solving eqs.~(3.13),(3.15) and (3.16), we can generate the multi-soliton 
 (multilump/ multidromion) solutions for the IE I. For example, we take 
 \begin{eqnarray}
 m_1(\xi,t)=-\frac {2\alpha_1X_2^2+2\alpha_2X_1^2+\frac {\alpha_1\alpha_2
 (\alpha_1+\alpha_2)^2}{2 (\alpha_1-\alpha_2)^2}}{(X_1 X_2-\frac {\alpha_1\alpha_2
 (\alpha_1+\alpha_2)^2}{4 (\alpha_1-\alpha_2)^2)^2}+\frac {1}{4}
 (\alpha_1X_2+\alpha_2X_1)^2},
 \end{eqnarray}
 where $X_i=\xi-\frac {2t}{\alpha_i}+\delta_i,\;i=1,2$, which describes the
 scattering of two lumps of the form given in eq. (4.1a) and $m_2(\eta,t)$ given
 in eq.~(4.1b). On solving eqs.~(3.13), (3.15) and (3.16), we can generate a (2,1)
 lump solution for IE I. Hence by appropriately choosing the boundaries
 $m_1(\xi,t)$ and $m_2(\eta,t)$, one can generate multidromions or multilumps
 and in general multisoliton structures.
 
 To conclude, in the case of IE also the Hirota method has proved to be a
 straightforward but powerful 
 tool to obtain different kinds of localized and other solutions. This has
 become feasible by appropriate bilinearization procedure in the light cone
 coordinates. We have also reported how the explicit forms of various localized
 solutions can be deduced. 

\ack

SV would like to thank Dr. S. Murugesh for useful discussions. The work of
ML forms the part of the Department of Science and Technology, Government of
India research project.


\section*{References}

\begin{thebibliography}{50}
\bibitem{laks:1976}
Lakshmanan M 1977 Phys. Lett. {\bf 61A} 53  
\bibitem{Ishi:1984}
Ishimori Y 1984 Prog. Theor. Phys. {\bf 72} 33
\bibitem{laks:1998} 
Lakshmanan M, Myrzakulov R, Vijayalakshmi S and Danlybaeva A K 1998 J. Math.
Phys. {\bf 39} 3765  
\bibitem{BGK:1990}
Konopelchenko B G and Matkarimov B T 1990 J. Math. Phys. {\bf 31} 2737
\bibitem{BGK:1991}
Dubrovsky V G and Konopelchenko B G 1991 Physica {\bf D48} 367
\bibitem{BGK:1993}
Konopelchenko B G 1993 {\it Solitons in Multidimensions} (Singapore: World
Scientific)
\bibitem{BGK:1992}
Dubrovsky V G and Konopelchenko B G 1992 Physica {\bf D55} 1
\bibitem{IMAI}
Imai K 1997 Prog. Theor. Phys. {\bf 98} 1013
\bibitem{laks:physica}
Lakshmanan M and Daniel M 1981 Physica {\bf 107A} 533 
\bibitem{hie}
Hietarinta J 1990 Phys. Lett. {\bf A149} 113
\bibitem{RR97a}
Radha R and Lakshmanan M 1997 Chaos, Solitons and Fractals {\bf 8} 17
\bibitem{RR97b}
Radha R and Lakshmanan M 1997 J. Math. Phys. {\bf 38} 292
\bibitem{RR97c}
Radha R and Lakshmanan M 1997 J. Phys. A:Math. Gen. {\bf 30} 3229
\bibitem{BGKstu}
Konopelchenko B G and Dubrovsky V G 1992 Stud. Appl. Math. {\bf 86} 219
\bibitem{Lou}
Tang X Y, Lou S Y and Zhang Y 2002 Phys. Rev. E {\bf 66} 046601
\end{thebibliography}
\end{document}